# Stability of the 5/2 Fractional Quantum Hall State in a Corbino Disc Sample with In-plane Electric Fields

Zheyi Zhu[1], Pengjie Wang[1], Hailong Fu[1], L. N. Pfeiffer[2], K. W. West[2], Rui-Rui Du[1,3], Xi Lin[1,3,*]

1 International Center for Quantum Materials, Peking University, Beijing 100871, China
2 Department of Electrical Engineering, Princeton University, Princeton, New Jersey 08544, USA
3 Collaborative Innovation Center of Quantum Matter, Beijing 100871, China

* xilin@pku.edu.cn

*Among the various geometries to study the fractional quantum Hall effect in two dimensional electron gas, the Corbino disc owns the advantage to probe the bulk properties directly. In this work we explore the influence of in-plane electric fields on the stability of the 5/2 fractional quantum Hall state realized in Corbino geometry. The effect of weak electric fields is investigated at ultra-low temperatures in order to compare with a theoretical proposal of enhanced Pfaffian state under weak electric fields.*

**Keywords:** fractional quantum Hall effect, 5/2 state, Corbino geometry, in-plane electric fields

## 1. Introduction

Since the discovery of the fractional quantum Hall (FQH) effect [1], around 100 FQH states have been observed and most of them are odd-denominator states [2]. In 1987, the first even denominator state, 5/2 FQH state was first observed with a longitudinal resistance minimum and a developing Hall resistance plateau [3], and its exact quantization was reported in 1999 with higher sample quality and lower electron temperature [4]. The even denominator state cannot directly share the theoretical explanation for the odd-denominator states. In the last 30 years, the 5/2 state has been drawing continuous attention both from theoretical and experimental aspects [5, 6]. A potential interest in the 5/2 state origins from that its elementary excitations might obey non-Abelian statistics, which makes this state a candidate for fault-tolerant topological quantum computation [7-10]. Numerous candidates with either Abelian or non-Abelian statistics have been proposed for the 5/2 state [8, 11-19]. The most theoretical plausible states are the non-Abelian Pfaffian state and its particle-hole conjugate, the anti-Pfaffian state [12-14]. There are also some other candidate states, such as the Abelian 331 state, which is the Halperin's generalization of Laughlin's wave function originally for a bilayer system [15, 16].

To identify the wave-function of the 5/2 state or the statistics of its quasi-particles, experiments have been carried out by measuring different properties of the 5/2 states [5, 6]. A common approach is to explore the 5/2 state from its edge current. For example, the quantum point contact (QPC) can be used to bring counter-propagating edge currents close to each other and induce quasi-particle tunneling. Shot noise measurements were carried out through QPC and have proved the e/4 charge of quasi-particles at the 5/2 state [20]. Quasi-particle tunneling experiments within a QPC could explore the statistics of the 5/2 state by probing the strength of Coulomb interaction between quasi-particles [21-24]. The results of the latest work [24] indicated that Abelian and non-Abelian states compete at filling factor 5/2. In interference experiments, e/4 and e/2 period interference oscillations supporting non-Abelian statistics have been observed [25, 26]. Experiments focused on the edge states have provided important information on the statistics of the 5/2 state, and further experiments to demonstrate the statistics or quasi-particle manipulation are still carrying on.

In addition to the edge, the bulk property also contribute information to identifying the statistics of the 5/2 state. The non-Abelian statistics can be revealed by the entropy carried by quasi-particles [8, 27-29], which is related to specific heat measurement at the 5/2 state [30]. Besides, other indirect methods have also been proposed aiming at the bulk properties, including the thermopower [27], temperature dependence of either the electrochemical potential or the orbital magnetization [28], and the anisotropy induced by in-plane magnetic field [31-33]. In the pursuit of bulk properties, the Corbino disc is an appropriate geometry to carry out measurements. The circular symmetry supports the isotropic transport in radial direction and excludes the effect of edge states, thus gives a direct probe to the bulk, which is superior over van der Pauw and Hall bar geometries. However, realization of fractional quantum Hall state in Corbino disc has been technically difficult and the successful attempt of the 5/2 state in Corbino disc is rare, with only one observation of the 5/2 state [34] and one specific heat measurement [30]. In the Hall bar geometry, the thermal power signal may be suppressed by the appearance of disorder, while in the Corbino geometry the thermal power signal from only quasi-particles could be measured [35]. Therefore, realization of the 5/2 state in Corbino geometry sample enriches the approach for investigation of quasi-particles' statistics [27, 35].

Recently, a theoretical work [36] studied the 5/2 Pfaffian state with in-plane electric fields in Corbino geometry. In that work, an enhancement of the Pfaffian state under sufficiently weak electric fields was predicted, while strong electric fields would destroy the state. Particularly in Corbino geometry an enhancement requires a certain polarity of the applied fields. The verification of this prediction will not only support the Pfaffian theoretical model and non-Abelian statistics, but also serve as a method to stabilize the 5/2 FQH state. Experimentally, this proposal can be realized by the ac differential measurement with a dc voltage bias applied across a Corbino disc.

In this work, we confirm the realization of the 5/2 FQH state in a Corbino device by the corresponding Hall plateau from a van der Pauw sample of the same wafer. Information of the 5/2 energy gap is obtained by varying temperature and electron density is obtained from SdH oscillations. We explore the effect of in-plane electric fields on the 5/2 FQH state at dilution refrigeration temperatures. Under small dc voltage bias, no enhancement of the 5/2 state is observed within our resolution at an excitation of 50 μV; under large dc voltage bias, the 5/2 state is destroyed by strong electric fields as predicted. The stability of the 5/2 FQH state under higher temperature is further explored.

## 2. Sample and Experiment Information
The Corbino device was fabricated from a GaAs/AlGaAs heterostructures with quantum well width 25 nm. The density of two-dimensional electron gas (2DEG) is $4.3 \times 10^{11}$ cm$^{-2}$ and the mobility is $1.2 \times 10^7$ cm$^2$ V$^{-1}$ s$^{-1}$. The Corbino disc owns an inner diameter of 1.8 mm and an outer diameter of 2.0 mm. In the Corbino geometry, pseudo four-terminal differential conductance was measured. As a result, only longitudinal transport with vanishing of conductance at a quantum Hall state is available and the plateau of Hall resistance is absent. We also fabricated a van der Pauw sample from the same wafer and measured together with the

Corbino sample, in order to confirm the filling factors.

The measurement setup of the Corbino sample is shown in Fig. 1(a), and that of the van der Pauw sample is shown in Fig. 1(b). The longitudinal conductance of the Corbino sample and the Hall resistance of the van der Pauw sample were measured in the same dilution fridge. The differential measurements were taken using a standard Lock-in technique at 17 Hz with a dc bias applied serving as an in-plane electric field. Between 10 and 100 μV, the transport data showed no difference during our test measurement and we used 50 μV for all the final measurements. The base temperature for this work was 30 mK. Before cooling down to the base temperature, the samples were illuminated by a red light-emitting diode at 4.5 K and 15 μA for 1 hour.

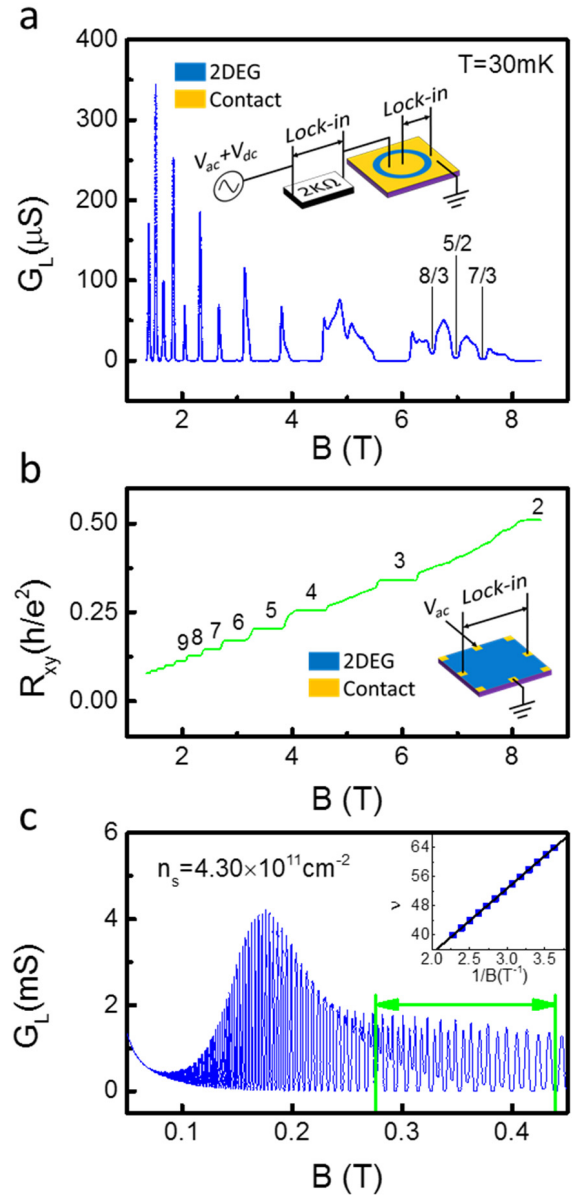

**FIG. 1.** Magnetic field dependence of the longitudinal differential conductance in Corbino sample and Hall resistance in van der Pauw sample. (a) Differential measurement setup and longitudinal differential conductance for the Corbino sample. (b) Differential measurement setup and Hall resistance for the van der Pauw sample. (c) SdH oscillation of the Corbino sample at low magnetic fields. The 2DEG density $n_s$ is

(4.30±0.01)×10$^{11}$ cm$^{-2}$, fitted within area indicated by green arrows, and the corresponding filling factors are labeled in the inset.

## 3. Results

The 2DEG density of the Corbino sample is determined from the SdH oscillations at low magnetic fields (Fig. 1(c)). As a result, the filling factors of the minimums in the longitudinal conductance of the Corbino sample can be labeled. In addition, the integer quantum Hall plateaus in the van der Pauw sample (Fig. 1(b)) coincide with the vanishing of Corbino longitudinal conductance (Fig. 1(a)), which also verified the filling factors we determined from the SdH oscillations. The filling factors of the well-formed minimums of conductance between filling factor 2 and 3 are calculated as 2.671±0.007, 2.540±0.006 and 2.371±0.006. The most likely FQH states should be 8/3, 5/2 and 7/3. The best quality part of the wafer is used for the Corbino sample, and the chip used for the van der Pauw sample is not as high quality as the Corbino sample, so the fractional plateaus from the van der Pauw sample in the second Landau levels are not fully developed.

The measurements of longitudinal conductance between the filling factor of 2 and 3 were carried out at different temperatures, as shown in Fig. 2(a). The temperature evolution of FQH states is observed with a deeper dip at a lower temperature. The center of the 5/2 state in magnetic field is determined from the dip of the longitudinal conductance at relatively high temperature, which is 6.992±0.005 T (Fig. 2(b)). From the relation $\sigma_{xx} = \sigma_0 e^{-\Delta/2k_BT}$ at 6.992T, the energy gap of the 5/2 FQH state can be calculated as 189 mK from the fitting of conductance versus reciprocal of temperature, as shown in Fig. 2(c).

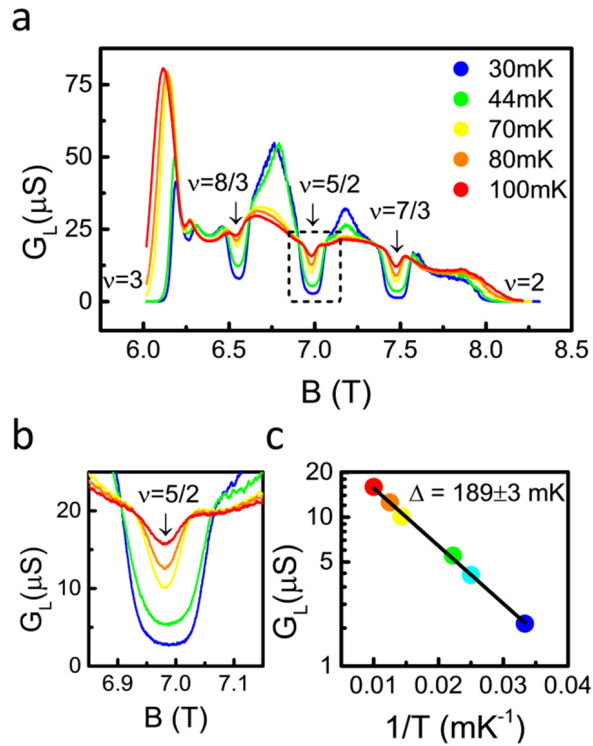

**FIG. 2.** The 5/2 FQH state in the Corbino sample. (a) Temperature evolution of the 5/2, 8/3 and 7/3 states, with a base temperature of 30 mK. (b) The zooming in around the filling factor of 5/2 from the dash square in

(a). (c) The fitting of conductance minimum of the 5/2 FQH state shows an energy gap of 189±3 mK.

The effect of weak in-plane electric fields on the 5/2 FQH state was explored at 32 mK and at the center of the 5/2 state in magnetic field. Fig. 3(a) exhibits the behavior of conductance under in-plane electric fields, which are applied through dc bias voltage. The voltage difference across the sample is converted to the electric field strength through dividing the voltage by the width of Corbino ring, considering the width (0.2 mm) is significantly smaller than the inner and outer diameter (1.9 mm on average). By applying both positive and negative dc voltages, electric fields are generated in both polarities. Within the resolution of measurement system, the conductance noise of about 0.1 μS and a sweep step of electric field strength of $2.4\times10^{-3}$ V/m, no decrease of longitudinal conductance was observed at the magnetic field center of the 5/2 state. Similar measurements were also carried out at different magnetic fields around the filling factor 5/2, and we did not observe a stability enhancement of the 5/2 state from the weak in-plane electric fields in this work.

The effect of strong in-plane electric fields on the longitudinal conductance at the base temperature is exhibited in Fig. 3(b) at different magnetic fields. Fig. 3(c) shows the data of white dash line in Fig. 3(b). In Fig. 3(c), the longitudinal differential conductance increases rapidly under sufficiently high fields and then decreases to a saturation value, which represents the breakdown of the 5/2 FQH state [37]. The 5/2 state becomes easier to destroy with the deviation from the center of the 5/2 FQH state, as expected. In the breakdown of integer quantum Hall states confined with a QPC [38], the states are more fragile at the high magnetic field side. Similar observation is found for the 5/2 state in this measurement although the asymmetry is not as distinct as the previous measurement.

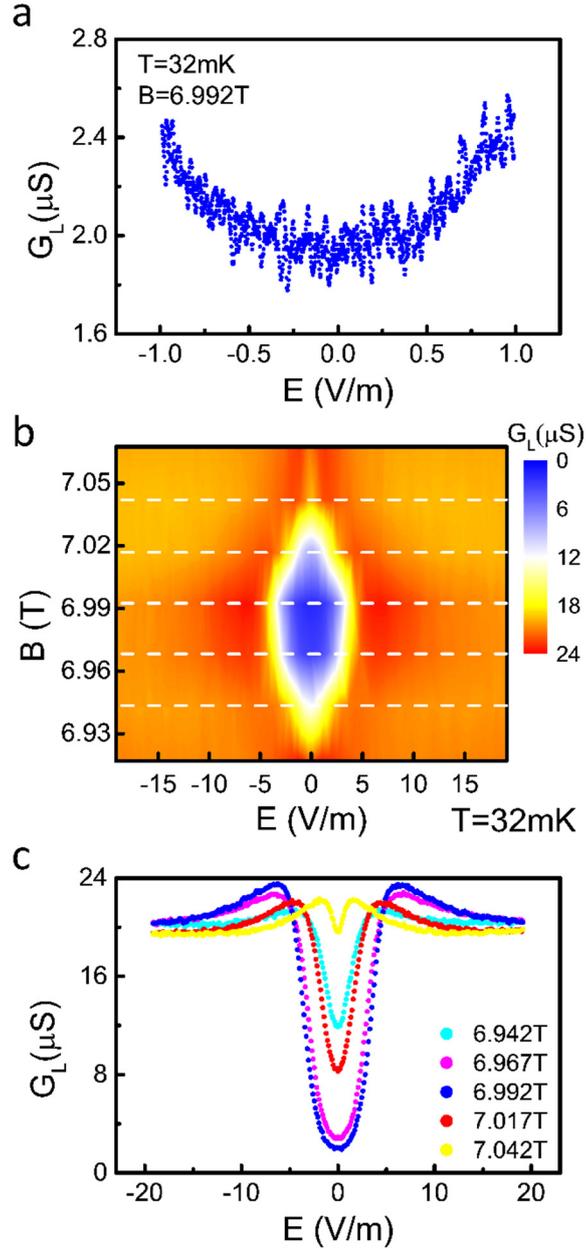

**FIG. 3.** The longitudinal conductance of the Corbino sample around ν=5/2 at 32 mK. (a) The longitudinal conductance at the center of the 5/2 FQH state under weak scanning electric field. (b) The longitudinal conductance as functions of the magnetic field and the electric field. (c) The cut scans of white dash lines in (b). The lowest dip comes from 6.992 T, the center of the 5/2 FQH state.

The stability of the 5/2 FQH state under different magnetic fields and temperatures are further explored, as shown in Fig. 4. The data are arranged in the order of magnetic fields ,which in Fig. 4(d) is nearest to the center of 5/2 state. As the magnetic field deviates from the center, similar to Fig. 3(b), the 5/2 state becomes less resistive to electric field, which can be observed under higher temperature. Besides, as a function of electric field, the differential conductance show a pair of shoulder on both polarity, and these shoulders become sharper with the deviation of magnetic field.

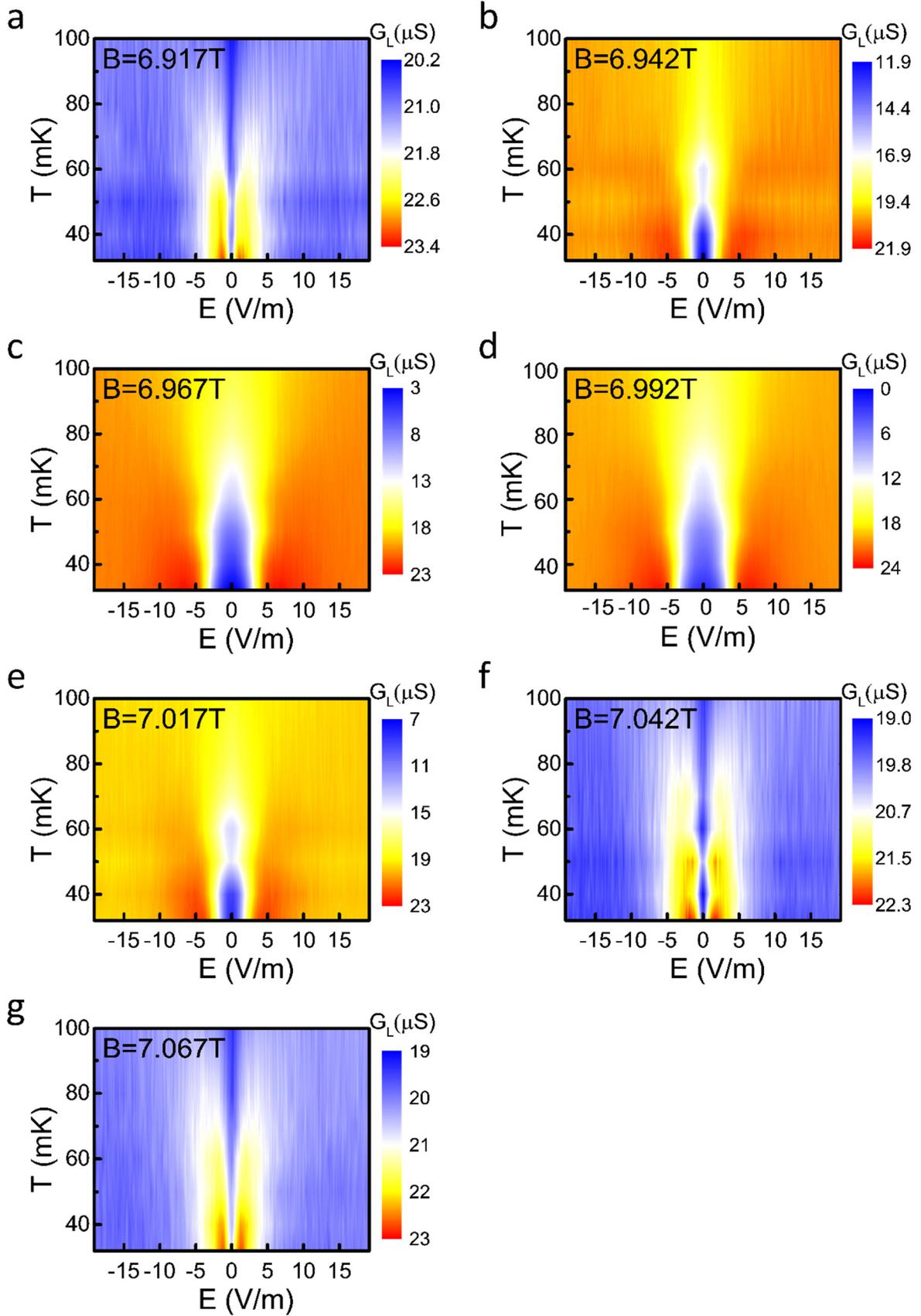

FIG. 4. The longitudinal conductance of the Corbino sample as functions of temperature and electric fields. From (a) to (g), the magnetic field increases at an interval of 0.025 T.

## 4. Discussions

The predicted enhancement of the 5/2 FQH state under weak in-plane electric field was not observed in our experiment with an excitation of 50 μV, which may result from the following reasons. Firstly, the electron temperature in this work may be not low enough to compare with the theory which assumes zero temperature. Secondly, the predicted enhancement might be weaker than the experimental signal to noise ratio, or the critical electric field corresponding to enhancement is smaller than our finest voltage scan step 0.5 μV. Thirdly, the experimental setup may not exactly represent the theoretical model. In the theoretical model, the Corbino disk's center ohmic contact is treated as a mathematical point, which means the electric fields are non-zero in most region of disk, while the central contact in this experiment has to occupy a finite area. Besides, the random disorder in a real sample and the unavoidable roughness of the ring boundary may cause the rotational asymmetry of the disc.

The inversion symmetry in radial direction is lack in Corbino geometry, and the effect of the electric field should depend on the field polarity as expected in [36]. However in this work, the stability of the 5/2 state exhibits same feature in both polarities, which may come from the large size of central contact and narrow width of Corbino ring. In a theoretical study on the 5/2 state with Corbino geometry [39], the phase diagram has been explored with respect to confining strength and Landau level mixing, and the stripe phase is proposed near the Pfaffian and anti-Pfaffian state. Since the electric field serves to rearrange the density distribution, a non-uniform stripe state is a potential candidate competing with the Pfaffian state. In addition, the width of the Corbino disc is 0.2 mm, 2 orders of magnitude larger than typical QPC where quasi-particle tunneling can be induced between counter-propagating edge currents, which should exclude the effects of edge states in this experiment.

For the temperature dependence of stability of the 5/2 state, in most cases it obeys the monotonic tendency as the conductance increases with more thermal excitation. However, in Fig. 4(b), 4(e) and 4(f), with the deviated magnetic field, the differential conductance at near-zero electric field exhibits a maximum at 50 mK as a function of temperature. Hence the stability of the 5/2 state under electrical field is not monotonic as a function of temperature, which argues that the electric fields play more roles in the stability of the 5/2 state than simply causing a heating effect.

## 5. Conclusion

In summary, a 5/2 FQH state in a Corbino disc is demonstrated experimentally, which enables further works such as thermopower measurement using the Corbino geometry. The filling factor is confirmed by the van der Pauw sample fabricated from the same wafer. The stability of the 5/2 FQH state is explored with in-plane electric fields. At weak electric fields, the theoretical prediction of enhancement of the 5/2 state is not observed with our experiment conditions.

**Acknowledgements:**
The work at PKU was funded by NSFC (Grant No. 11674009, 11274020 and 11322435) and NBRPC (Grant No. 2015CB921101). The work at Princeton University was funded by the Gordon and Betty Moore

Foundation through the EPiQS initiative Grant GBMF4420, by the National Science Foundation MRSEC Grant DMR-1420541, and by the Keck Foundation.